\documentclass[proceedings, preprint]{rmaa}

\usepackage{paralist}

\usepackage[utf8]{inputenc}
\usepackage{psfrag,color}
\newcommand{\msun}{M$_\odot$}
  
\SetYear{2010}
\SetConfTitle{XIII Latin American Regional IAU Meeting}

\title{Disruption of Dwarf Satellite Galaxies without Dark Matter} 

\author{
  R. A. Casas\altaffilmark{1} 
  and P. Kroupa\altaffilmark{2}}

\altaffiltext{1}{Departamento de Física, Universidad Nacional de Colombia, Ciudad Universitaria, Bogotá, Colombia (racasasm@unal.edu.co).} 

\altaffiltext{2}{Institut für Astronomie, Universität Bonn, Auf dem
  Hügel 71, D-53121 Bonn, Germany (kroupa@astro.uni-bonn.de).} 

\shortauthor{Casas \& Kroupa}
\shorttitle{Dwarf Satellites without Dark Matter}

\listofauthors{R. A. Casas \& P. Kroupa}
\indexauthor{Casas, R. A.}
\indexauthor{Kroupa, P.}

\abstract{
The evolution of a satellite galaxy of a Milky Way like galaxy has been
studied using N-Body simulations. The initial satellites, containing
$10^6$ particles, have been simulated by a Plummer sphere, while the potential
of the host galaxy is a three component rigid potential: disc, bulge and dark matter halo. It has been found that several orbits of the satellites allow for the existence, for about 1
Gyr or more, of an out-of-equilibrium body that could be
interpreted as a dSph satellite galaxy of the Milky Way. In addition, from the study of the evolution of the
mass-to-light ratios of satellites that show a disrupted phase it has been found that it is
possible that some dSph galaxies of the Milky Way with large M/L ratios might
not be dark matter dominated and that their high mass to light ratios are
observed because they are out of equilibrium objects.
}

\resumen{
La evolución de galaxias satélite de la Vía Láctea es estudiada mediante simulaciones numéricas de N-cuerpos.
Los satélites iniciales, que contienen $10^6$ partículas fueron modelados
mediante esferas de Plummer, mientras que el potencial de la galaxia nodriza
corresponde a un potencial rígido de tres componentes: disco, esferoide y halo de materia oscura.
Se encontró que un conjunto de órbitas del
satélite permite la existencia por más de $10^9$ años de remanentes fuera del
equilibrio que podrían ser interpretados como galaxias dSph de la Vía
Láctea.  Adicionalmente, mediante el estudio de la evolución de la razón masa luminosidad de los satetélites se encontró que  es posible que algunas galaxias dSph de la Vía Láctea con razones M/L altas pudiesen no estar dominadas por materia oscura y que dichos valores altos se deban a que se trata de objetos fuera del equilibrio.
}

\addkeyword{galaxies: dwarf}
\addkeyword{galaxies: evolution}
\addkeyword{galaxies: formation}
\addkeyword{galaxies: interactions}

\begin{document}
\maketitle

\section{Introduction}
Dwarf spheroidal galaxies (dSph) are stellar ellipsoidal, gas poor and with
low surface brightness systems.  For some dSph satellites mass-to-light (M/L)
ratios as large as a few hundred are inferred. A M/L ratio larger than 10 is
usually associated with a dark matter dominated system. Nevertheless, an
alternative possibility to explain large M/L ratios relying on Newtonian
physics, is that some of the observed dSph satellites may not be in virial
equilibrium. That such systems may exist is shown by the simulations of the
long-term evolution of initially spherical satellite galaxies with $3 \times
10^5$ particles and typical masses of $10^7 \ M\odot$
\citep{1997NewA....2..139K}. 

In this paper we present the evolution of a couple of
 initially in equilibrium spherical satellites with $10^6$ particles orbiting a galaxy that resembles the Milky Way.

\section{N-body Simulations}

The evolution of the satellites orbiting the Milky Way has been simulated
using the tree-code feature of the Gadget-2 N-body code
\citep{2005MNRAS.364.1105S}.  Each simulation has been run for 12 Gyr and the
evolution of some quantities of the satellites like their Lagrange radii, number of bound particles, central surface brightness, half light
radius and the M/L ratio have been studied. The last three quantities have been obtained
projecting the remnant object as it would appear to an observer on earth. 

The host galaxy has been modeled using a three component rigid potential: a
Miyamoto--Nagai potential for the disk, a Hernquist potential for the bulge
and  a Logarithmic potential for the dark matter halo. The parameters of the
model are the ones appearing in \citet{1995ApJ...451..598J}. 
 
The initial satellites are modeled as Plummer spheres with $10^6$ particles,
plummer radius of 0.3 kpc and cuttof radius of 1.5 kpc. Their initial masses
are $10^7$ and $10^8$ M$_{\odot}$ respectively, and intrinsically dark matter
free. For the construction of the spheres the algorithm proposed by
\citet{1974A&A....37..183A} has been used.

\section{Satellite Evolution}

\begin{figure}[!h]
\includegraphics[width=\linewidth,clip]{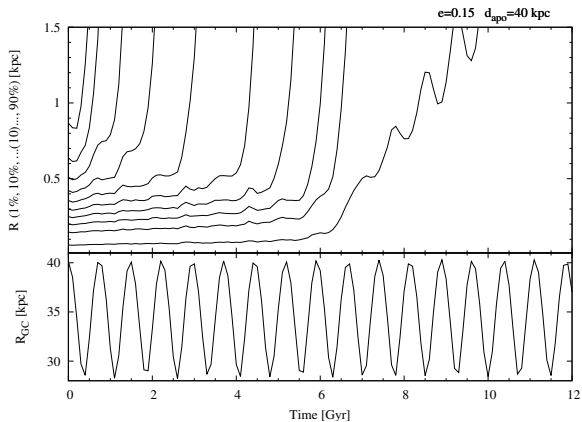}
\caption{Lagrange radii containing (bottom to top) 1\%, 10\%, 20\%, ... , 90\% of the initial mass of a satellite with $10^7$ \msun{}. Bottom panel: galactocentric distance of the density maximum. The parameters of the orbit are given in the upper right corner}
\label{fig:lagrange-radii}
 \end{figure}

Satellites in very eccentric orbits are completely depopulated in a very short
time interval, usually in a single perigalactic passage. On the other hand, satellites on orbits with very low
eccentricity can retain more than 90\% of their initial mass over a Hubble
time. These kind of orbits lead to an object that does not reach an
out-of-equibrium state. The intermediate case corresponds to satellites that
loose more than 90\% of their initial mass but retain more 
than 1\% of the mass for about 1 Gyr or more (see Figure \ref{fig:lagrange-radii}). These satellites present an out-of-equilibrium
phase and are the objects we are more interested in.

\section{Mass-to-light ratios}

\begin{figure}[!htb]
\includegraphics[width=\linewidth,clip]{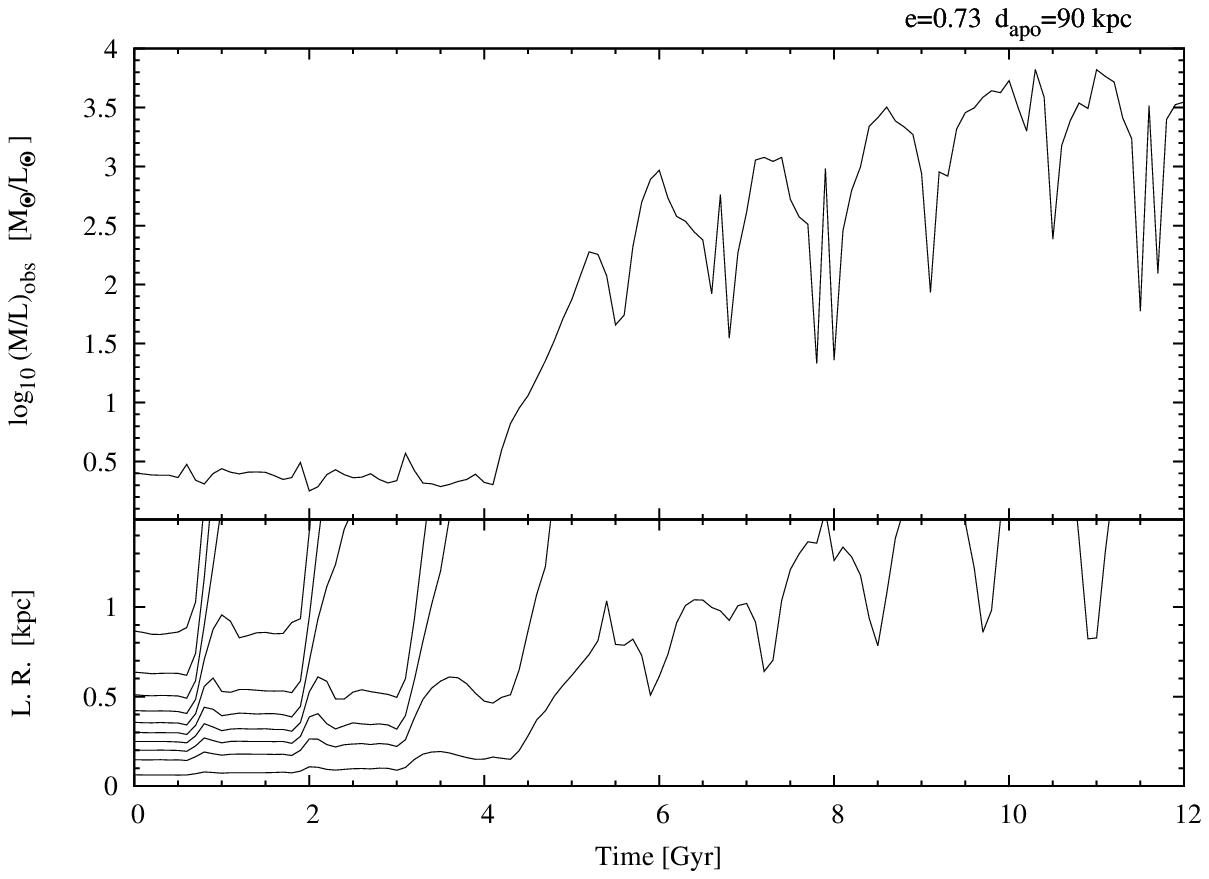}\\
\includegraphics[width=\linewidth,clip]{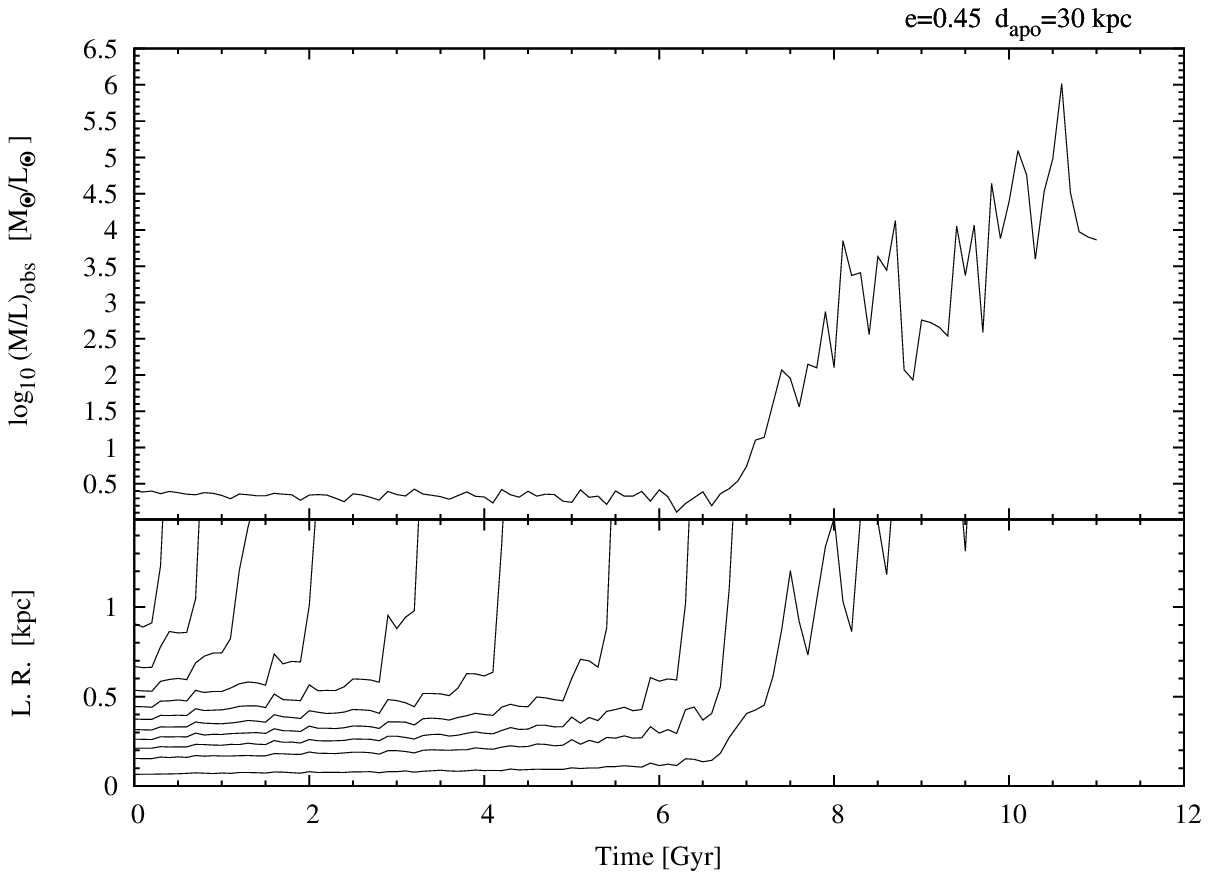}
\caption {Mass-to-light ratios for satellites with initial mass of $10^7 M_{\odot}$ (top) and $10^8 M_{\odot}$ (bottom). For both satellites the Lagrange radii, as in figure \ref{fig:lagrange-radii}, have been included to locate start of disrupted phase.}
\label{fig:m2l}
\end{figure}

In Figure \ref{fig:m2l} we plot the evolution of the M/L ratio for two
out-of-equilibrium satellites obtained using the
core fitting formula \citep[and references therein]{1986AJ.....92...72R}. The M/L ratio remains relatively constant
as the satellite is being smoothly depopulated. After the
satellite enters the disrupted phase the M/L ratio rises considerably,
reaching values of hundred and more. 
The values of the M/L ratios  of the satellites during the
disrupted phase show large oscillations. Thus, depending on the time of 
observation a satellite might show very different values of the M/L ratio. Values much larger than the intrinsic M/L ratio of the satellites can only be obtained during the disrupted phase.

\section{Concluding Remarks}

It is possible to obtain disrupted remnants
of initially spherical bound objects, that survive as out-of-equilibrium
systems for times longer than 1 Gyr, confirming and extending the results from
\citet{1997NewA....2..139K}.  The disrupted satellites show a high
M/L ratio, similar to that of some dSph galaxies. Those remnants are
candidates to be interpreted by an observer on earth as dSph-like galaxies. 
Hence, it is possible that some dSph galaxies of the Milky Way with large M/L ratios might
not be dark matter dominated and that their high M/L ratios are
observed because they are out of equilibrium systems.

\section*{Aknowledgements}
R.A.C. aknowledges financial support  from the \textit{ Convocatoria Nacional
  de Investigación 2008 de la Universidad Nacional de Colombia}.

\bibliography{my_citations}
\end{document}